\documentclass[a4paper,twocolumn,aps,nofootinbib,preprintnumbers,superscriptaddress,amsmath,amssymb,amsfonts]{revtex4}
\usepackage{graphicx}
\usepackage{amsmath}
\usepackage[latin1]{inputenc}
\usepackage[english]{babel}
\usepackage[T1]{fontenc}
\usepackage{amssymb}
\usepackage{amsfonts}
\usepackage{epsfig}
\usepackage{colordvi}
\usepackage{psfrag}
\usepackage{color}
\usepackage{dcolumn}
\usepackage{multirow}
\usepackage{hyperref}
\usepackage{epstopdf}
\usepackage{times}
\hypersetup{
  colorlinks=true,        
  linkcolor=blue,         
  citecolor=magenta,      
}

\newcommand{\be}{\begin{eqnarray}}
\newcommand{\ee}{\end{eqnarray}}

\renewcommand{\d}{\mbox{${\rm d}$}}

\begin{document}

\title{Orbits in bootstrapped Newtonian gravity}

\author{Anna~D'Addio}
\email{anna.daddio@unina.it}
 \affiliation{Dipartimento di Fisica ``E.~Pancini'', Universit\`a di Napoli ``Federico II'', Complesso Universitario di Monte Sant'Angelo, via Cinthia Edificio~6, 80126~Napoli, Italy}
 \affiliation{%
  I.N.F.N., Sezione di Napoli,Complesso Universitario di Monte Sant'Angelo, via Cinthia Edificio~6, 80126~Napoli, Italy
}
\author{Roberto Casadio}%
 \email{casadio@bo.infn.it}
\affiliation{%
  Dipartimento di Fisica e Astronomia ``A.~Righi'', Universit\`a di Bologna,via Irnerio~46, 40126~Bologna, Italy
}%
\affiliation{%
 I.N.F.N., Sezione di Bologna, I.S.~FLAG,viale B.~Pichat~6/2, 40127~Bologna, Italy.
}%

\author{Andrea~Giusti}%
 \email{agiusti@phys.ethz.ch}
\affiliation{%
  Institute for Theoretical Physics, ETH Zurich,Wolfgang Pauli Strasse~27, 8093~Zurich, Switzerland.
}%

\author{Mariafelicia~De~Laurentis}%
 \email{mariafelicia.delaurentis@unina.it}
 \affiliation{Dipartimento di Fisica ``E.~Pancini'', Universit\`a di Napoli ``Federico II'', Complesso Universitario di Monte Sant'Angelo, via Cinthia Edificio~6, 80126~Napoli, Italy}
\affiliation{%
  I.N.F.N., Sezione di Napoli,Complesso Universitario di Monte Sant'Angelo, via Cinthia Edificio~6, 80126~Napoli, Italy
}%
\affiliation{%
  Laboratory of Theoretical Cosmology, Tomsk State University,634050~Tomsk, Russia.
}%
\date{\today}

\begin{abstract}
Bootstrapped Newtonian gravity is a nonlinear version of Newton's law which 
can be lifted to a fully geometric theory of gravity starting from a modified potential. 
Here, we study geodesics in the bootstrapped Newtonian effective metric in {\em vacuum} 
and obtain bounds on a free parameter from Solar System data and S-star orbits near our Galaxy center. 
These bounds make {\em vacuum} bootstrapped Newtonian gravity experimentally 
indistinguishable from General Relativity.
\end{abstract}

\maketitle

\section{Introduction}
\label{s:intro}
\setcounter{equation}{0}
General Relativity is presently the most successful theory for describing the gravitational interaction
at the classical level.
Its own failure is marked by the prediction of the formation of geodesic singularities whenever a trapped surface arises from the gravitational collapse of a compact object.\footnote{We
also recall that pointlike sources are mathematically incompatible with the Einstein field equations~\cite{geroch}.}
Such considerations open up the possibility that significant departures from General Relativity
might occur where our experimental data do not yet place strong enough constraints,
like for example in regions of strong gravity near a very massive source.
However, Einstein's field equations are not linear and this makes it difficult to modify the laws of gravity in the
strong-field regime without affecting also the weak-field behavior, since these regimes are likely to be related
nontrivially in any nonlinear theories.
\par
The bootstrapped Newtonian gravity~\cite{BootN,Casadio:2020mch} is an attempt at
investigating these issues in a somewhat simplified context.
The approach, based on Deser's conjecture~\cite{deser}, consists of retrieving the full Einstein's theory including gravitational self-coupling terms in the Fierz-Pauli action.\footnote{This
idea is indeed older, see {\em e.g.}~Ref.~\cite{Feynman}.} 
These additional terms must be consistent with diffeomorphism invariance, in order to preserve the covariance of any (modified) metric theory.
We can obtain different modified gravitational theories depending on the choice of boundary conditions in the reconstruction procedure~\cite{rubio}.
A key observation is that a practically effective dynamics can be derived only starting with a ``small'' contribution of matter sources.
For large astrophysical sources, this implies that the matter source must also be included in
a nonperturbative way.
In the present approach this task is addressed starting from the Fierz-Pauli action corresponding to the potential generated by an arbitrarly large static source, and putting in extra terms representing gravitational self-coupling.
Furthermore, the coupling constants for the additional terms are not fixed to their Einstein-Hilbert values in order to accommodate for diverse underlying dynamics.
This approach then results in a nonlinear equation including pressure effects and the gravitational self-interaction terms to next-to-leading order in the Newton constant, whose solution is the gravitational potential operating on test particles at rest.
Such equation was useful to investigate compact objects~\cite{Casadio:2019cux,Casadio:2020kbc,Casadio:2019pli} and coherent quantum states~\cite{Casadio:2016zpl,Casadio:2017cdv}.\footnote{These quantum states show some of the
properties~\cite{ciarfella} found in the corpuscular model of black holes~\cite{DvaliGomez}.
However, we shall not discuss quantum aspects in this work.}
\par
 The motion of (test) particles and photons in the surrondings of a compact object represents the most immediate tool to gather information on the gravitational potential in which they revolve.
In Ref.~\cite{Casadio:2021gdf}, a full (effective) metric tensor was obtained from the bootstrapped
Newtonian potential, which allows one to study these trajectories in general, and to compare them with
results from General Relativity.
The requirement that the resulting theory of gravity is covariant is satisfied by the use of an effective metric tensor, since the bootstrapped Newtonian dynamics is implicitly assumed to be invariant after coordinate transformations.
Nonetheless, the particular metric found in Ref.~\cite{Casadio:2021gdf} differs from the Schwarzschild 
geometry; hence, it is not a solution of the Einstein equations in the vacuum.
An effective fluid is therefore present, as was already noted in the cosmological context~\cite{cosmo}.
\par
The bootstrapped effective metric is given as a function of parameterized
post-Newtonian (PPN) parameters~\cite{weinberg} in the weak-field expansion.
These parameters can be consistently chosen so as to minimize deviations from the Schwarzschild metric
only up to a point.
In fact, some of the PPN parameters are uniquely related, and at the PPN order determined in
Ref.~\cite{Casadio:2021gdf}, they can be expressed in terms of one free parameter. 
In this work, we report on a phenomenological investigation aiming at placing bounds on 
this remaining free parameter from the measured precessions in the Solar System ~\cite{DeMartino2018,Moyer200,Will2018}and from the study
of S-star orbits around the black hole in the center of the Galaxy ~\cite{Eckart1996,Eckart1997,Gillessen:2009,Gillessen2009L,Ghez1998,Ghez:2008}.
\par
The paper is organized as follows.
In Sec.~\ref{s:boot}, we briefly review the equation for the bootstrapped Newtonian 
potential and its solution in the vacuum.
We then just recall the full effective metric reconstructed from this potential, which is 
then used to analyze Solar System data and S-star motions in Sec.~\ref{s:app}. We conclude with comments and an outlook in Sec.~\ref{s:conc}.
\section{Bootstrapped Newtonian vacuum}
\label{s:boot}
\setcounter{equation}{0}
We shall only review briefly the derivation of the bootstrapped Newtonian equation, since all the
details can be found in Refs.~\cite{Casadio:2017cdv,BootN,Casadio:2019cux,Casadio:2019pli}. We shall use units with
the speed of light $c=1$ in this section.
We start from the Lagrangian for the Newtonian potential $V=V(r)$ generated by a static
and spherically symmetric source of density $\rho=\rho(r)$, to wit
\begin{equation}
L_{\rm N}[V]
=
-4\,\pi
\int_0^\infty
r^2\,\d r
\left[
\frac{\left(V'\right)^2}{8\,\pi\,G_N}
+V \rho
\right]
\ .
\label{LVn}
\end{equation}
The corresponding Euler-Lagrange field equation is given by Poisson's
\begin{equation}
\dfrac{1}{r^2}\,\dfrac{d}{d r}
\left(r^2\,\frac{dV}{dr}\right)=
4\,\pi\,G_N\,\rho
\ ,
\label{poisson}
\end{equation}
where we recall that the radial coordinate $r$
is the one obtained from harmonic coordinates~\cite{weinberg,Casadio:2021gdf}.
We next couple $V$ to a gravitational current proportional to its own energy density,
\begin{equation}
J_V
\simeq
4\,\frac{d U_{ N}}{d \mathcal{V}} 
=
-\dfrac{\left[V'(r)\right]^2}{2\,\pi\,G_N}
\ ,
\end{equation}
where $\mathcal{V}$ is the spatial volume and $U_{\rm N}$ is the Newtonian potential energy.
We also add the ``one loop term'' $J_{\rho}\simeq-2\,V^2$, which couples to $\rho$, and
the pressure term $p$~\cite{Casadio:2019cux}.
The total Lagrangian then reads 
\begin{align}
L[V]
=&
-4\,\pi
\int_0^\infty
r^2\,\d r
\left[
\frac{\left(V'\right)^2}{8\,\pi\,G_N}
\left(1-4\,q_V\,V\right)
\right.\nonumber\\&\left.+\left(\rho+3\,q_p\,p\right)V
\left(1-2\,q_\rho\,V\right)\right]\,
\label{LagrV}
\end{align}
where the coupling constants $q_V$, $q_p$ and $q_\rho$ can be used to track the effects of the different
contributions.
For instance, the case $q_V=q_p=q_\rho=1$ reproduces the Einstein-Hilbert action at next-to-leading order
in perturbations around Minkowski~\cite{Casadio:2017cdv,Casadio:2019cux,Casadio:2019pli}.
Finally, the bootstrapped Newtonian field equation reads
\begin{eqnarray}
\dfrac{1}{r^2}\,\dfrac{d}{d r}
\left(r^2\,\dfrac{d V}{d r}\right)
&&=
4\,\pi\,G_N
\dfrac{1-4\,q_\rho\,V}{1-4\,q_V\,V}
\left(\rho+3\,q_p\,p\right)
\nonumber\\&&+\dfrac{2\,q_V\left(V'\right)^2}
{1-4\,q_V\,V}
\ ,
\label{EOMV}
\end{eqnarray}
which must be solved along with the conservation equation $p' = -V'\left(\rho+p\right)$. 
\subsection{Vacuum potential}
\begin{figure}[!t]
    \centering
    \includegraphics[scale=0.9]{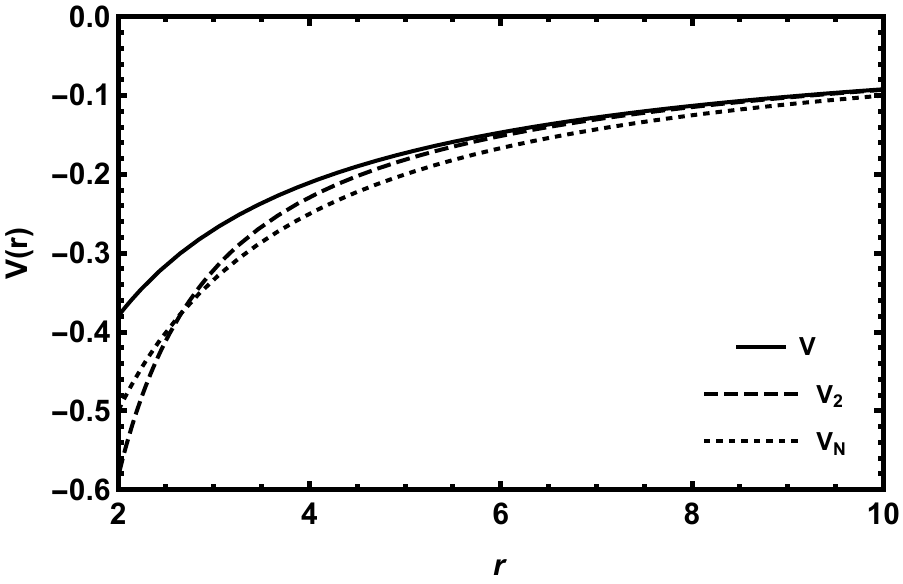}
    \caption{Bootstrapped Newtonian potential $V$ in Eq.~\eqref{potential} compared to
    its expansion $V_2$ from Eq.~\eqref{Vasy} and to the Newtonian potential $V_{\rm N}$
    (for $q=1$).}
    \label{f:V}
\end{figure}

In vacuum, we have $\rho=p=0$, and Eq.~\eqref{EOMV} simplifies to
\begin{equation}
\frac{1}{r^2}\,\frac{d}{d r}
\left(r^2\,\frac{d V}{d r}\right)
=
\frac{2\,q \left(V'\right)^2}{1-4\,q\,V}
\ ,
\label{EOMV0}
\end{equation}
where we renamed $q\equiv q_V$ for simplicity.
The exact solution was found in Ref.~\cite{BootN} and reads
\begin{eqnarray}
\label{potential}
V(r)
=
\frac{1}{4\,q}
\left[1-\left(1+\frac{6\,q\,G_N\,M}{r}\right)^{2/3}\right]
\ .
\end{eqnarray}
The asymptotic expansion away from the source yields
\begin{equation}
V_{2}
\simeq
-\frac{G_N\,M}{r}
+q\,\frac{G_N^2\,M^2}{r^2}
-q^2\,\frac{8\,G_N^3\,M^3}{3\,r^3}
\ ,
\label{Vasy}
\end{equation}
so that the Newtonian behavior is always recovered (for $q=0$) and the post-Newtonian terms
are seen to depend on the coupling $q$ (see Fig.~\ref{f:V}).
\
\
\subsection{Vacuum effective metric}
\label{ss:metric}
A complete spacetime metric was reconstructed from the vacuum potential~\eqref{potential}
in Ref.~\cite{Casadio:2021gdf}.
The procedure is rather cumbersome, and we shall only recall here a few main steps leading
to the necessary expressions in the weak-field regime. We explicitly show the speed of light
$c$ from here on.
One starts from the PPN form~\cite{weinberg}
\begin{widetext}
\begin{equation}
ds^2\simeq\left[1-\alpha\dfrac{2R_g}{\bar{r}}+(\beta-\alpha\gamma)\dfrac{2R_g^2}{\bar{r}^2}+(\zeta-1)\dfrac{8R_g^3}{\bar{r}^3}\right]c^2dt^2 +\left[1+\gamma\dfrac{2R_g}{\bar{r}}+\xi\dfrac{4R_g^2}{\bar{r}^2}+\sigma\dfrac{8R_g^3}{\bar{r}^3}\right]d\bar{r}^2+\bar{r}^2d\Omega^2
\end{equation}
\end{widetext}
where $R_g=\frac{G_N\,M}{c^2}$ and $\bar{r}$ is the areal radius, which differs from the radial coordinate $r$ in which
the potential~\eqref{potential}
is expressed.
The latter is obtained from harmonic coordinates and the two radial coordinates are
related by~\cite{Casadio:2021gdf}
\begin{eqnarray}
r
\simeq&&
\bar{r}
+
(1-3\,\gamma)\,\frac{R_g}{2}
+\nonumber\\
&&+(1-3\,\gamma+2\,\gamma^2-2\,\Xi)\,
\frac{R_g^2}{\bar{r}}
\ ,
\end{eqnarray}
in which $\Xi$ is a free parameter.
Furthermore, we have
\begin{equation}
q
=
\beta+\frac{\gamma-1}{2}
\ .
\label{eq:q}
\end{equation}
\par
We can next set $\alpha=1$ by simply absorbing this coefficient in the definition of the mass $M$~\cite{adm},
and  $\beta=\gamma=1$ in order to satisfy the experimental constraints $|\gamma-1|\simeq|\beta-1|\ll 1$.
From Eq.~\eqref{eq:q}, this is tantamount to setting $q=1$, as expected.
The higher order PPN parameters are then fully determined by $\Xi$ according to 
\begin{eqnarray}
\xi
&\!\!=\!\!&
1+\Xi
\label{eq:xi}
\\
\zeta
&\!\!=\!\!&
1-\frac{5+6\,\Xi}{12}
=
\frac{13-6\,\xi}
{12}
\label{eq:zeta}
\\
\sigma
&\!\!=\!\!&
\frac{9+14\, \Xi}
{4}
\ .
\label{eq:sigma}
\end{eqnarray}
As already noted in Ref.~\cite{Casadio:2021gdf}, the General Relativistic PPN combination $\xi=\zeta=1$
cannot be obtained for any value of $\Xi$, and the bootstrapped metric for which we have the minimum deviation
from the Schwarzschild form is thus given by 

\begin{eqnarray}
ds^2
\simeq&&
-\left[
1
-\frac{2\,R_g}{r}
-(5+6\,\Xi)\,\frac{2\,R_g^3}{3\,c^6\,r^3}
\right]
c^2\,dt^2
\nonumber
\\
&&
+
\left[
1
+
\frac{2\,R_g}{r}
+
(1+\Xi)\,\frac{4\,R_g^2}{r^2}
\right.\nonumber\\&&\left.
+
(9+14\, \Xi)\frac{2\,R_g^3}{r^3}
\right]
dr^2
+
r^2\,d\Omega^2
\ ,
\label{eq:g}
\end{eqnarray}
in which we drop the bar from the areal coordinate for simplicity from now on.
We can see that there are contributions in the metric coefficients which cannot be reduced to the
Schwarzschild expressions.
This deviation from the Schwarzschild solution is encoded by the free parameter $\Xi$,
whose value is \textit{a priori} unknown and must be constrained by observations.
In particular, we will test these corrections by analyzing the 
planets in the Solar System and S-stars motion around Sgr~A*.
The geodesic equations
\begin{equation}
\Ddot{x}^{\mu}+\Gamma^{\mu}_{\alpha\beta}\,\dot{x}^{\alpha}\,\dot{x}^{\beta}
=
0,
\end{equation}
where a dot indicates the derivative with respect to the proper time, can be equivalently computed using the Euler-Lagrange equations
\begin{equation}
\dfrac{d}{d s}
\dfrac{\partial\mathcal{L}}{\partial\dot{x}^{\mu}}
-\dfrac{\partial\mathcal{L}}{\partial x^{\mu}}
=
0
\ , 
\end{equation}
with $\mathcal{L}=g_{\alpha\beta}\,\dot x^\alpha\,\dot x^\beta=-1$ for a massive object.
From the metric in Eq.~\eqref{eq:g}, one then finds  
\par
\begin{widetext}
\begin{align}
\Ddot{r}
&
=
\frac{R_g
\left\{
4\, (1+\Xi)\, R_g\,r\,\dot{r}^2
+R_g^2
\left[3
\left(
9+14\Xi
\right)
\dot{r}^2
-c^2
\left(5+6\Xi\right)
\dot{t}^2
\right]
+r^2\left(\dot{r}^2-c^2\,\dot{t}^2\right)
\right\}
+r^5\,(\dot{\theta}^2+\dot{\phi}^2\,\sin^2\theta)}
{r
\left[2\, (9+14\, \Xi)\, R_g^3+4\,(1+\Xi)\, R_g^2\, r+2\,R_g\,r^2+\,r^3
\right]}
\label{eq:ddr}
\\
\Ddot{\theta}
&=
\dot{\phi}^2\,\sin\theta\,\cos\theta
-\frac{2\,\dot{r}\,\dot{\theta}}{r}
\\
\Ddot{\phi}
&=
-\frac{2\,\dot{\phi}}{r}\,(\dot{r}+r\,\dot{\theta}\,\cot\theta)
\\
\Ddot{t}
&
=
\frac{6\,\dot{r}\,\dot{t}
\left[
(5+6\,\Xi)\, R_g^3
+R_g\,r^2
\right]}
{2\,(5+6\,\Xi)\, R_g^3\,r
+6\,R_g\,r^3
-3\,r^4}
\ .
\label{eq:ddt}
\end{align}
\end{widetext}
 The third and fourth equations are the usual conservation equations for the angular momentum
and energy conjugated to $t$, respectively.
Spherical symmetry as usual implies that the orbital motion occurs on a plane which we can
arbitrarily set at $\theta=0=\dot\theta$.


The above parametric system of nonlinear differential equations can be integrated numerically
in order to study the orbits.

\subsubsection{Precession}
\label{ss:precession}
%
%
\par
It is easy to express the perihelion precession in terms of the PPN parameters~\cite{weinberg}.
At leading order, one has
\begin{equation}
\label{eq:eq1}
\Delta\phi^{(1)}
=
2\,\pi
\left(2-\beta+2\,\gamma\right)
\dfrac{R_g}{\ell}
\ ,
\end{equation}
where $\ell=a\,(1-e^2)$ is the \textit{semilatus rectum}, $a$ is the semimajor axis
and $e$ is the eccentricity.
For $\beta=\gamma=1$, Eq.~(\ref{eq:eq1}) reproduces the General Relativistic result
\begin{equation}
\label{eq:eq2}
\Delta\phi_S^{(1)}
=
6\,\pi\,\frac{R_g}{\ell}
\ .
\end{equation}
The second order correction depends on $\xi$ and $\zeta$, and for $\beta=\gamma=1$,
it reads~\cite{Casadio:2021gdf}
\begin{align}
\Delta\phi^{(2)}
&
=
\pi\left[(41+10\xi-24\,\zeta)
+
(16\,\xi-13)\,\frac{e^2}{2}\right]
\frac{R_g^2}{\ell^2}
\nonumber
\\
&
\simeq
\pi\left[(37+22\,\Xi)+(3+16\,\Xi)\,\frac{e^2}{2}\right]
\frac{R_g^2}{\ell^2}
\nonumber
\\
&
\simeq
\Delta\phi_S^{(2)}
+
2\,\pi\left[
11\,\xi-7+4\,(\xi-1)\,e^2\right]
\frac{R_g^2}{\ell^2}
\ ,
\end{align}
where the General Relativistic result $\Delta\phi_S^{(2)}$ corresponds to $\xi=\zeta=1$.
From Eqs.~\eqref{eq:xi} and \eqref{eq:zeta}, it follows that we cannot have $\xi=\zeta=1$ for any value of $\Xi$,
and a deviation from General Relativity remains. 
\section{Astronomical tests}
\label{s:app}
\setcounter{equation}{0}
In order to constrain the free parameter of the bootstrapped Newtonian potential, $\Xi$,
we confronted the theoretical results exposed in Sec.~\ref{ss:metric} with astronomical 
data.

To infer a range of validity for $\Xi$, we compared the analytical expression of the precession with the observed values of the perihelion advance of Solar System's planets (Sec.~\ref{ss:precession2}).

Then, we turned our attention to the Galactic Center, and we studied the motion of S-stars
orbiting around Sgr A*.
To constrain $\Xi$, we let it vary in a given range and fit the corresponding simulated
orbits to astrometric observations.
In particular, we adopted a fully relativistic approach which consists of integrating numerically
Eqs.~\eqref{eq:ddr}-\eqref{eq:ddt} in order to get the mock orbits, instead of solving Newton's
law with the standard potential replaced by the modified one. 
\subsection{Perihelion precession in the Solar System}
\label{ss:precession2}
In order to constrain $\Xi$ we can start from the Solar System planets whose orbital precession
has been measured, namely Mercury, Venus, Earth, Mars, Jupiter and Saturn~\cite{2015MNRAS.451.3034N}.
The confidence region for $\Xi$ can be identified as the set of values such that the precession
\begin{equation}
\Delta\phi=\Delta\phi^{(1)}+\Delta\phi^{(2)}
\label{eq:eq3}
\end{equation}
is compatible with the observations.
The planetary parameters\footnote{The reported values are taken from NASA fact sheet at
https://nssdc.gsfc.nasa.gov/planetary/factsheet/.},
the corresponding observed values of the precession~\cite{2015MNRAS.451.3034N} and
the General Relativistic value obtained by Eq.~(\ref{eq:eq2}) are reported in Table~\ref{tab:tab1}
from first to seventh columns.
\begin{table*}
    \centering
    \begin{tabular}{cccccccc}
    \hline\hline
    Planet & $a(10^6km)$ & $P(years)$ & $i(^{\circ})$ & $e$ & $\Delta\phi_{obs}(''/cy)$ & $\Delta\phi_S(''/cy)$ & $[\Xi_{min};\Xi_{max}]$ \\
    \hline
    $\textbf{Mercury}$&$57.909$&$0.24$&$7.005$&$0.2056$&$43.1000\pm0.5000$&$42.9822$&$[-89708.7;144995]$\\
    $\textbf{Venus}$&$108.209$&$0.61$&$3.395$&$0.0067$&$8.6247\pm0.0005$&$8.6247$&$[-1149.67;1167.47]$\\
    $\textbf{Earth}$&$149.596$&$1.00$&$0.000$&$0.0167$&$3.8387\pm0.0004$&$3.83881$&$[-3660.86;2094.96]$\\
    $\textbf{Mars}$&$227.923$&$1.88$&$1.851$&$0.0935$&$1.3565\pm0.0004$&$1.35106$&$[155248.;179879.]$\\
    $\textbf{Jupiter}$&$778.570$&$11.86$&$1.305$&$0.0489$&$0.6000\pm0.3000$&$0.0623142$&$[5.46709\times10^8;1.92679\times10^9]$\\
    $\textbf{Saturn}$&$1433.529$&$29.45$&$2.485$&$0.0565$&$0.0105\pm0.0050$&$0.0136394$&$[-1.57315\times10^8;3.59618\times10^7]$\\
    \hline\hline
    \end{tabular}
\caption{Values of semimajor axis ($a$), orbital period ($P$), tilt angle ($i$), eccentricity ($e$), observed orbital precession
($\Delta\phi_{obs}$), orbital precession as predicted by General Relativity ($\Delta\phi_S$) and constraints on $\Xi$ for Solar System's planets.}
    \label{tab:tab1}
\end{table*}
\begin{figure*}[!ht]
    \centering
    \includegraphics[scale=0.6]{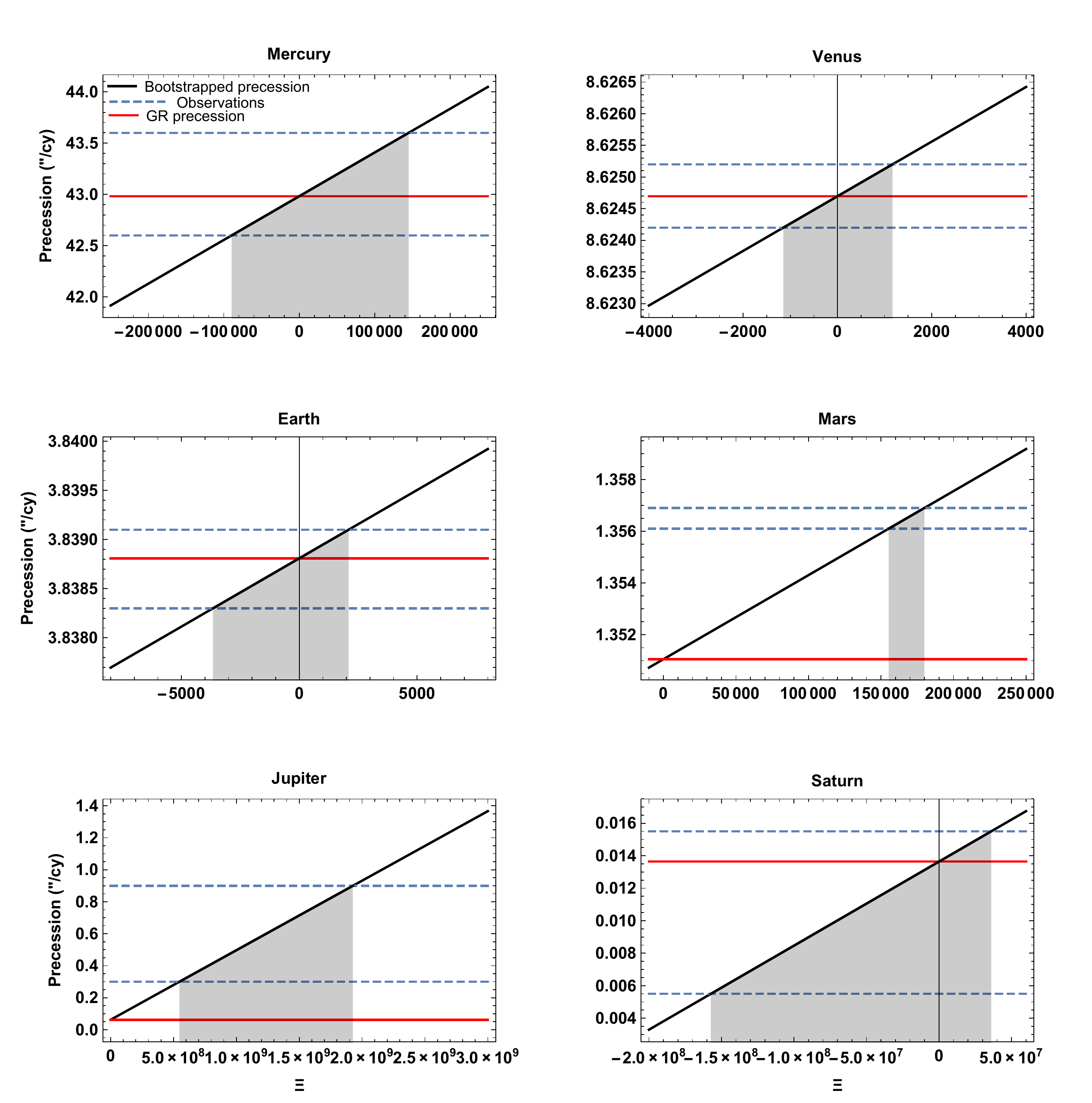}
    \caption{Bootstrapped orbital precession as a function of the parameter $\Xi$.
    Black lines give the theoretical prediction from Eq.~\eqref{eq:eq3}, blue dashed lines represent
    the measurements adapted from Ref.~\cite{2015MNRAS.451.3034N} and red lines depict the General Relativistic
    values as in Eq.~\eqref{eq:eq2}.
    Confidence regions for $\Xi$ are shaded in gray.}
    \label{fig:fig1}
\end{figure*}
The allowed region of $\Xi$ for each planet is defined as the range of values compatible with data, having as extremes the
values of $\Xi$ solving the equation
\begin{equation}
\Delta\phi=\Delta\phi_{obs}
\ .
\end{equation}
The inferred lower and upper limits on $\Xi$ are reported in the last column of Table~\ref{tab:tab1}, and the included area
is depicted in Fig.~\ref{fig:fig1} for each planet (gray shades).
It is worth noticing the discrepancy between the General Relativistic value (the red line) and the observed precession
(blue dashed lines) for Mars and Jupiter; it could be attributed to the incomplete subtraction of nonrelativistic effects
from the observed value, complicated by the presence of the asteroid belt between Mars and Jupiter, and the presence
of an anomalous residual precession \cite{2015MNRAS.451.3034N,2013MNRAS.432.3431P}.
\par
The tightest interval on the parameter $\Xi$ is obtained with Venus, for which it can vary between $-1149.67$ and $1167.47$.
We can use the values defining such an interval to predict the precession for Uranus, Neptune and Pluto,
for which no observation is available.
The results, summarized in Table~\ref{tab:tab2}, show that the bootstrapped theory predictions are in perfect agreement
with General Relativity. 
\begin{table*}
    \centering
    \begin{tabular}{ccccccc}
    \hline\hline
    Planet & $a(10^6km)$ & $P(years)$ & $i(^{\circ})$ & $e$ &  $\Delta\phi_{S}(''/cy)$ & $[\Delta\phi_{min};\Delta\phi_{max}]$ \\
    \hline
    $\textbf{Uranus}$&$2872.463$&$84.01$&$0.772$&$0.0457$&$0.00238404$&$[0.00238404;0.00238405]$\\
    $\textbf{Neptune}$&$4495.060$&$164.786$&$1.769$&$0.0113$&$0.000775374$&$[0.000775373;0.000775375]$\\
    $\textbf{Pluto}$&$5869.656$&$247.936$&$17.16$&$0.2444$&$0.000419669$&$[0.000419669;0.00041967]$\\
    \hline\hline
    \end{tabular}
    \caption{Orbital parameters from Nasa Fact Sheet, the General Relativistic prediction for the precession in the sixth column
    and the values predicted by the bounds on the parameter $\Xi$ of the bootstrapped theory deduced for Venus (see Table~\ref{tab:tab1}).}
    \label{tab:tab2}
\end{table*}
\par
Now it is useful to move to a different scale and analyze $S2$ (see Table \ref{tab4}), the only one among the S-stars whose precession was observed \cite{GRAVITY:2020gka}.
We can next calculate the precession for Mars, Jupiter, and $S2$ with the values of $\Xi$ as obtained by Mercury, Venus, Earth
and Saturn to check agreement with the corresponding Schwarzschild value and with the observations (Table \ref{tab3}). The results confirm the compatibility of our predictions with General Relativity.
The mean value of the parameter $\Xi$ such that
\begin{equation}
    \Delta\phi
    =
    \Delta\phi_S
\end{equation}
is given by
\begin{equation}
\Xi
=
-1.64236\pm 0.10305
\ .
\label{eq:c2}
\end{equation}
\begin{table*}
    \centering
    \begin{tabular}{cccccccc}
    \hline\hline
    Star & $a(AU)$ & $P(years)$ & $i(^{\circ})$ & $e$ & $\Delta\phi_{obs}(''/\text{orbit})$ & $\Delta\phi_S(''/\text{orbit})$ & $[\Xi_{min};\Xi_{max}]$ \\
    \hline
    $\textbf{S2}$&$1031.32$&$16.0455$&$134.567$&$0.884649$&$730.382\times(1.10\pm0.19)$&$730.382$&$[-103.066;326.398]$\\

    \hline\hline
    \end{tabular}
    \caption{For the star $S2$, orbital parameters~\cite{GRAVITY:2020gka}, observed orbital precession ($\Delta\phi_{obs}$),
    orbital precession as predicted by General Relativity ($\Delta\phi_S$), and constraints on $\Xi$.}
    \label{tab4}
\end{table*}
\begin{table*}
    \centering
    \begin{tabular}{cccccc}
    \hline\hline
   Object  & $\Delta\phi_S$ & $\Delta\phi(\Xi_{Mercury})$ &$\Delta\phi(\Xi_{Venus})$&$\Delta\phi(\Xi_{Earth})$&$\Delta\phi(\Xi_{Saturn})$\\
    \hline
    $\textbf{Mars}$&$1.35106$&$[1.34814;1.35577]$&$[1.35102;1.3511]$&$[1.35094;1.35113]$&$[-3.75855;2.5191]$\\
    $\textbf{Jupiter}$&$0.0623142$&$[0.0622752;0.0623773]$&$[0.0623137;0.0623147]$&$[0.0623126;0.0623151]$&$[-0.00607962;0.0779489]$\\
    $\textbf{S2}$&$730.382$&$[-57243.9;94435.7]$&$[-11.7295;1485.75]$&$[-1634.61;2085.15]$&$[-1.01666*10^8;2.32414*10^7]$\\
    \hline\hline
    \end{tabular}
    \caption{Precession for Mars, Jupiter, and $S2$ as predicted by confidence regions for $\Xi$ inferred from Mercury, Venus, Earth and Saturn.}
    \label{tab3}
\end{table*}

\subsection{S-star dynamics}
\label{ss:dynamic}
We can confirm the bounds on $\Xi$ deduced from orbital precessions by comparing them with results deduced from the analysis
of stellar orbits at the Galactic Center.
This further analysis consists in comparing simulated orbits in bootstrapped Newtonian gravity, obtained by integrating numerically
Eqs.~\eqref{eq:ddr}-\eqref{eq:ddt}, with observed orbits of three S-stars constructed by astrometric observations
(see Sec.~\ref{sec:data}).
In particular, we focused on stars $S2$, $S38$ and $S55$ for two main reasons:
among the brightest stars they are those with the shortest period.
These properties are desired because highly bright stars are less prone to being confused with other sources,
and a short period allows us to observe a larger part of the orbit in a given observation session.
For simplicity, we neglected perturbations from other members of the cluster and any extended matter structures. 
\subsubsection{Astrometric data}
\label{sec:data}
Astrometric data are taken from Ref.~\cite{Gillessen:2017}~\footnote{Data are publicly available on the electronic version of
Ref.~\cite{Gillessen:2017} at the link https://iopscience.iop.org/article/10.3847/1538-4357/aa5c41/meta.}
and cover $25$ years of observations performed in the near-infrared (NIR), where the interstellar extinction amounts to
about three magnitudes.
Different instruments have been used, which we briefly describe below.
\begin{enumerate}
\item
SHARP.- First high-resolution data of the Galactic Center were obtained in $1992$ with the SHARP camera at the European Southern Observatory's
(ESO) $3.5\,$m New Technology Telescope (NTT) in Chile, operating in Speckle mode with exposure times of $0.3\,$s, $0.5\,$s and $1.0\,$s.
The data, described in detail in Ref.~\cite{Schodel:2003gy}, led to the detection of high proper motion near the central massive object.
\item
NACO.- The first Adaptive Optics (AO) imaging data were produced by Naos-Conica (NACO) system,
mounted at the telescope Yepun ($8.0\,$m) of the VLT and starting to operate in $2002$.
It followed a great improvement due to larger telescope aperture, and the higher Strehl ratio (about $40\%$).
\item
GEMINI.- The dataset includes observations obtained by the $8\,$m telescope Gemini North in Mauna Kea, Hawaii.
These images, obtained using the AO system in combination with the NIR camera Quirc, were processed by the Gemini team.
\end{enumerate}
The astrometric calibration of these data, treated in Ref.~\cite{Gillessen:2009ht}, consists in the following steps:
obtaining high-quality maps of the S-stars, extracting pixel positions, and transforming them to a common astrometric
coordinate system.
In particular, the astrometric reference frame is implemented relating the S-stars positions to a set of selected reference
stars, which are in turn related to a set of Silicon Monoxide (SiO) maser stars whose positions relative to Sgr~A* is known.
\subsubsection{Fitting procedure}
The first step of the fitting procedure is the numerical integration of the system of parametric nonlinear
differential equations~\eqref{eq:ddr}-\eqref{eq:ddt} to produce stellar simulated orbits in bootstrapped
Newtonian gravity.
\begin{table*}
\begin{centering}
    \begin{tabular}{cccc}
    \hline\hline
Parameter           & S2                     & S38               & S55               \\ \hline
$a$ (mas)           & $125.058\pm0.041$      & $141.6\pm0.2$     & $107.8\pm1.0$     \\
$\Omega$ ($^\circ$) & $228.171\pm0.031$      & $101.06\pm0.24$   & $325.5\pm4.0$     \\
$e$                 & $0.884649\pm0.000066$  & $0.8201\pm0.0007$ & $0.7209\pm0.0077$ \\
$i$ ($^\circ$)      & $134.567\pm0.033$     & $171.1\pm2.1$     & $150.1\pm2.2$     \\
$\omega$ ($^\circ$) & $66.263\pm0.031$       & $17.99\pm0.25$    & $331.5\pm3.9$     \\
$t_p$ (yr)          & $2018.37900\pm0.00016$ & $2003.19\pm0.01$  & $2009.34\pm0.04$  \\
$T$ (yr)            & $16.0455\pm0.0013$     & $19.2\pm0.02$     & $12.80\pm0.11$  \\ 
$m_K$     &
13.95          &
17.       &
17.5     \\
Ref.          &
\cite{GRAVITY:2020gka} &
\cite{Gillessen:2017}              &
\cite{Gillessen:2017} \\
\hline\hline
    \end{tabular}
        \caption{Orbital parameters of $S2$, $S38$, and $S55$:
        semimajor axis  $a$, eccentricity $e$, inclination $i$, angle of the line of node $\Omega$,
        angle from ascending node to pericenter $\omega$, orbital period $T$, and the time of the
        pericenter passage $t_p$. }
        \label{tab1}
        \end{centering}
\end{table*}
\begin{table*}
\begin{centering}
    \begin{tabular}{cccc}
    \hline\hline
     Star&$M(M_{\odot})$&$R(kpc)$&Ref.\\
    \hline
    S2&$(4.261\pm0.012)\times10^6$&$8.2467\pm0.0093$&GRAVITY Collaboration \cite{GRAVITY:2020gka} \\
    S38&$(4.35\pm0.13)\times10^6$&$8.33\pm0.12$&Gillessen et al. \cite{Gillessen:2017}\\
    S55&$(4.35\pm0.13)\times10^6$&$8.33\pm0.12$&Gillessen et al. \cite{Gillessen:2017}\\
    \hline\hline
    \end{tabular}
        \caption{Parameters of the central BH: the mass $M$ and the distance $R$.} \label{tab2}
        \end{centering}
\end{table*}
\par
Preliminarily, we fix the Keplerian elements and the parameters of the central mass to the values reported
in Tables~\ref{tab1} and \ref{tab2}.
In particular, for the study of $S2$, we used the values obtained by the GRAVITY Collaboration~\cite{GRAVITY:2020gka},
and for $S38$ and $S55$, we used those obtained in Ref.~\cite{Gillessen:2017}.
In order to have a well-defined Cauchy problem, we must provide initial conditions for the four-dimensional coordinates
and their derivatives with respect to the proper time: \{$r(0)$, $\dot{r}(0)$, $\theta(0)$, $\dot{\theta}(0)$, $\phi(0)$,
$\dot{\phi}(0)$, $t(0)$, $\dot{t}(0)$\}. 
We assume that the star initially lies on the equatorial plane of the reference system, for which $\theta(0)=\pi/2$,
and that its initial velocity is parallel to the equatorial plane, that is $\dot{\theta}(0)=0$.
It then follows that $\ddot{\theta}(0)=0$ identically.
In particular, we set the initial conditions for $r$ and $\phi$ at the time of passage of the apocenter,
when the Cartesian coordinates of the star expressed in the orbital plane are given by
\begin{equation}
(x_{orb},y_{orb})
=
\left(-a\,(1+e),0\right)
\end{equation}
and the Cartesian components of its velocity read
\begin{equation}
(v_{x,orb},v_{y,orb})
=
\left(0,\dfrac{2\,\pi\,a^2}{T\,r}\,\sqrt{1-e^2}\right)
\ .
\end{equation}
The initial condition for $\dot{t}$ can be retrieved from the normalization of four-velocities requiring
that the geodesic is timelike.
\par
Starting from the initial conditions of each star, we proceed with an explicit Runge-Kutta numerical integration
of the relativistic equations of motion.
The results are the stars mock orbit in the orbital plane, described by a four-dimensional array
$\{t(\tau),r(\tau),\theta(\tau),\phi(\tau)\}$.
To compare the theoretical orbits with those observed from the Earth, we must project any point
$(x_{\rm orb}, y_{\rm orb})$ on the orbital plane into the point $(x, y)$ on the observer's sky plane.
Such a transformation is realized by applying the Thiele-Innes formulas~\cite{1930MNRAS,aitken}:
\begin{align}
    x&=l_1\,x_{\rm orb}+l_2\,y_{\rm orb}
    \\
    y&=m_1\,x_{\rm orb}+m_2\,y_{\rm orb}
    \ .
\end{align}
The Thiele-Innes elements $l_1$, $l_2$, $m_1$ and $m_2$ depend on the Keplerian elements by according to
\begin{align}
    l_1
    &=
    \cos\Omega\,\cos\omega-\sin\Omega\, \sin\omega\, \cos i
    \\
    l_2
    &=
    -\cos \Omega\, \sin\omega-\sin\Omega\, \cos\omega \cos i
    \\
    m_1
    &=
    \sin\Omega\, \cos\omega+\cos\Omega\, \sin\omega \cos i
    \\
    m_2
    &=
    -\sin\Omega\,\sin\omega+\cos\Omega\, \cos\omega\, \cos i
     .
     \end{align}
\par     
The second step consists in the fitting procedure itself, and has the aim to constrain the parameter $\Xi$.
Guided by the results obtained from the precession in Sec.~\ref{ss:precession}, we let it vary freely in an
appropriate range including the value~\eqref{eq:c2}.
For each value of $\Xi$ we repeated the aforementioned procedure to get the true positions $(x_i,y_i)$
and velocities $(\dot{x}_i,\dot{y}_i)$ of the stars at all the observed epochs.
After transforming the true positions into the apparent positions $(x_i^{th},y_i^{th})$, we computed the
reduced-$\chi^2$ distribution to quantify the discrepancy between theory and observations as
\begin{equation}
  \chi^2_{\rm red}
  =
  \frac{1}{2\,N-1}\,
  \sum_i^N\left[\left(\frac{x_i^{\rm obs}-x_i^{\rm th}}{\sigma_{x_i^{\rm obs}}}\right)^2
  +\left(\frac{y_i^{\rm obs}-y_i^{\rm th}}{\sigma_{y_i^{\rm obs}}}\right)^2
  \right]
  \ ,
\end{equation}
where $(x_i^{\rm obs},y_i^{\rm obs})$ and $(x_i^{\rm th},y_i^{\rm th})$ are respectively the observed and
the predicted positions, $N$ is the number of observations and $(\sigma_{x_i^{\rm obs}},\sigma_{y_i^{\rm obs}})$
are the observative uncertainties.
Finally, we calculated the likelihood probability distribution, $2\,\log\mathcal{L}=-\chi^2_{\rm red}(\Xi)$.
The best-fit value for $\Xi$ was derived as the point that maximizes the likelihood distribution.
\subsubsection{Results}
\begin{table}
    \centering
    \begin{tabular}{cc}
    \hline\hline
     Star&$\Xi$\\
    \hline
    $S2$&$-5900_{-44964.9}^{+39358.8}$\\
    $S38$&$25500_{-23312.88}^{+22607.1}$\\
    $S55$&$60400_{-87446.9}^{+81386}$\\
    Multi-star&$17400_{-32244.3}^{+30555.6}$\\
    \hline\hline
    \end{tabular}
    \caption{Best-fit values for $\Xi$.}
    \label{tab5}
\end{table}
\begin{figure*}[!ht]
    \centering
    \includegraphics[scale=0.45]{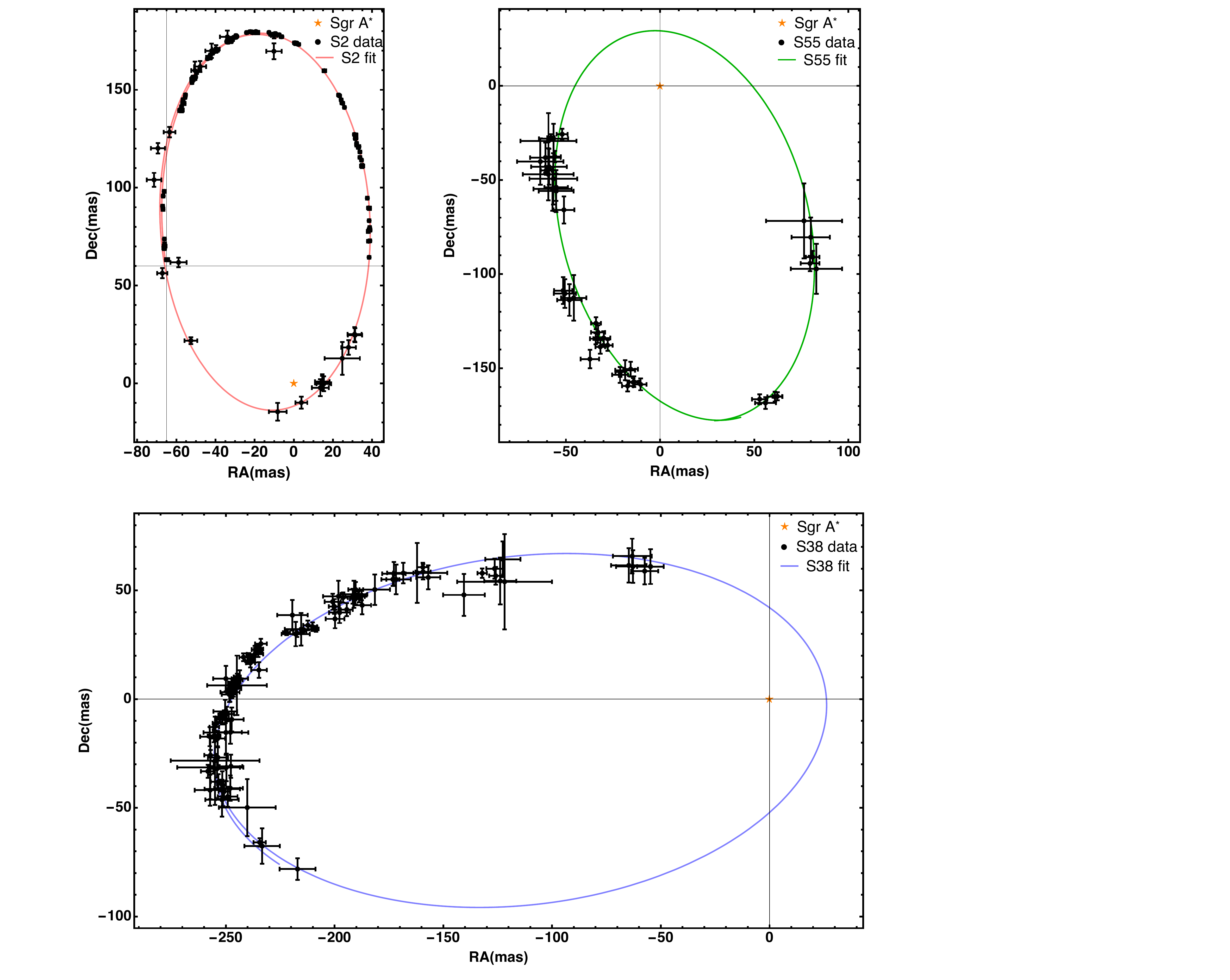}
    \caption{Comparisons between the NTT/VLT astrometric observations with their errors (black circles)
    and the theoretical best-fit orbits using parameters reported in the first three rows of Table~\ref{tab5}.
    The results for $S2$, $S55$ and $S38$ are illustrated respectively in the top left, top right, and bottom panels.}
    \label{fig:fig3}
\end{figure*}
\begin{figure*}[!ht]
    \centering
    \includegraphics[scale=0.6]{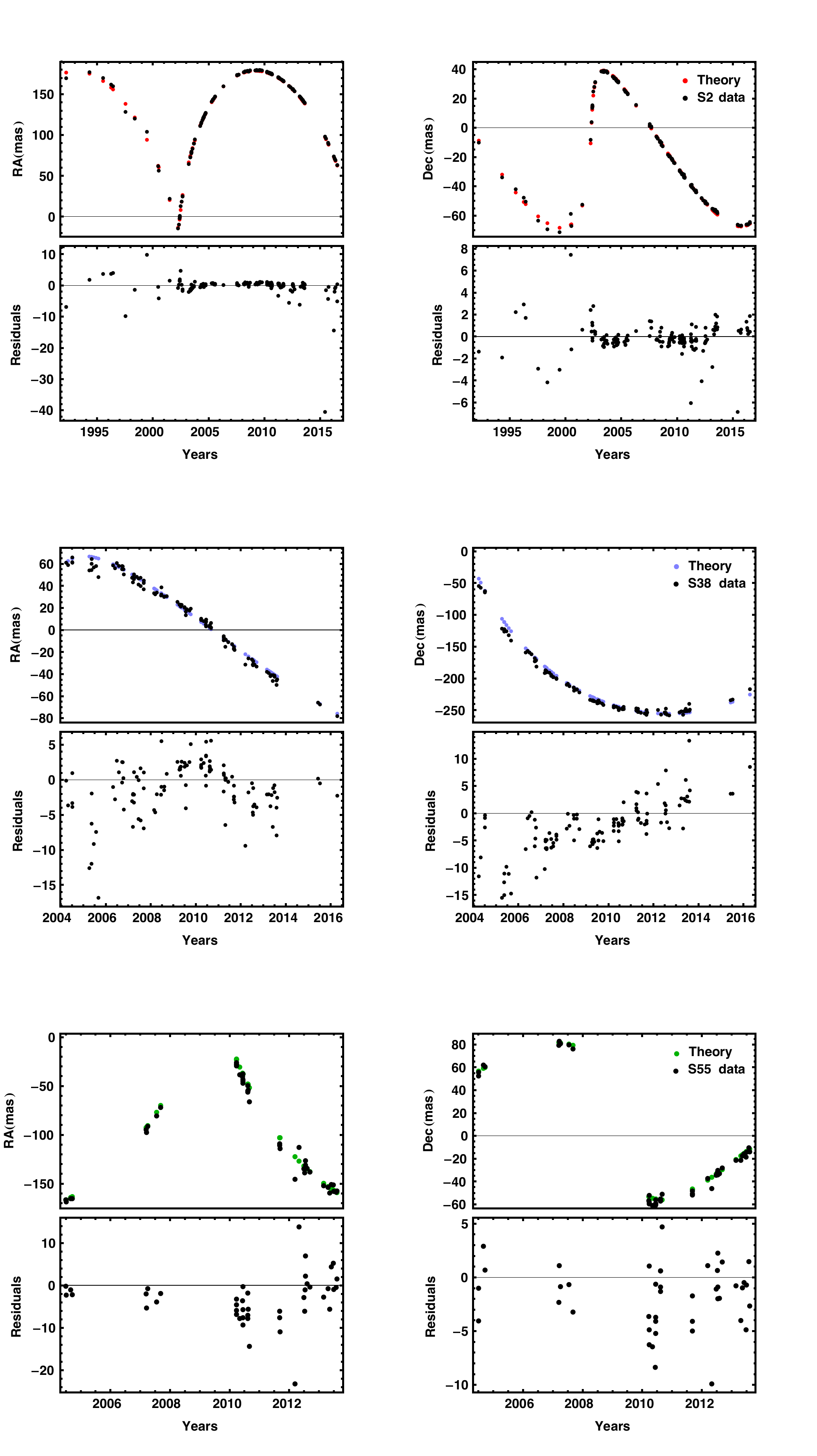}
    \caption{Top panels show the comparison between the observed and fitted coordinates, and
    bottom panels show the corresponding (O-C) residuals for $S2$, $S38$ and $S55$.}
    \label{fig:fig4}
\end{figure*}
\begin{figure*}[!ht]
    \centering
    \includegraphics[scale=0.7]{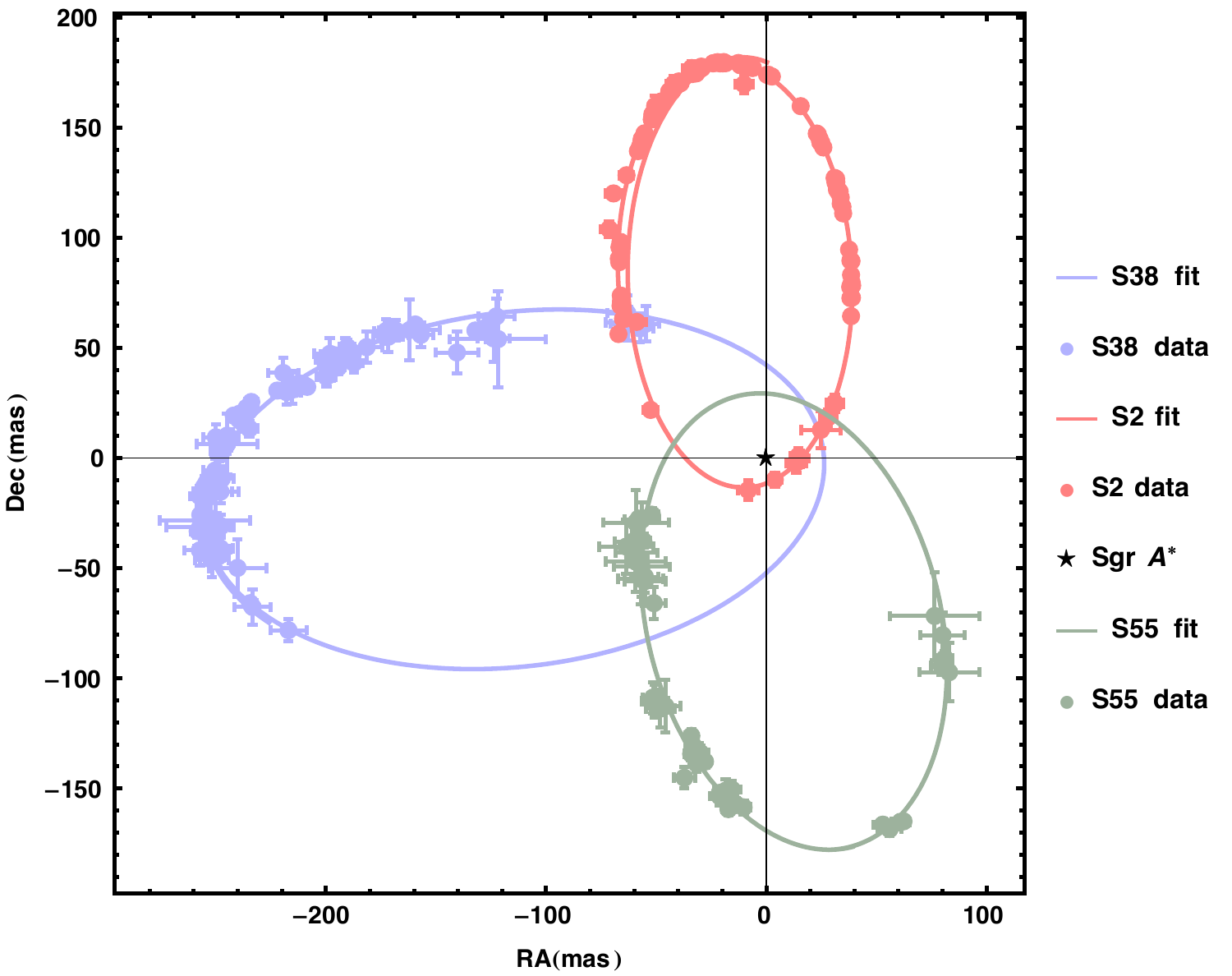}
    \caption{Best relativistic multistar orbit fit of $S2$, $S38$ and $S55$.}
    \label{fig:fig5}
\end{figure*}
Our results are summarized in Table~\ref{tab5} and represented in Figs.~\ref{fig:fig3}, \ref{fig:fig4} and \ref{fig:fig5}.
\par
In Fig.~\ref{fig:fig3} we show the comparison between best fit and observed orbits of the selected stars:
the top left panel, the top right panel, and the bottom panel illustrate the results respectively for $S2$, $S55$ and $S38$.
Astrometric data are reported with their own error bars to note the effectiveness of our fitting procedure.
\par
Figure~\ref{fig:fig4} depicts the comparisons between the observed and simulated coordinates with the corresponding residuals.
The left column contains the right ascension (RA), while the right column reports the declination (Dec).
It is worth noticing that in all stars and for both coordinates, residuals are larger at the beginning observing epochs,
and decrease as astrometric accuracy improves.
\par
Finally, we show in Fig.~\ref{fig:fig5} the orbits of the studied S-stars corresponding to the best multistar fit for
$\Xi=17400_{-32244.3}^{+30555.6}$ (last row of Table~\ref{tab5}).
As expected, the parameter $\Xi$ is compatible with the the mean value~\eqref{eq:c2} such that the bootstrapped
Newtonian precession recovers General Relativity.
\section{Conclusions}
\label{s:conc}
\setcounter{equation}{0}
In this paper we tested astronomically the bootstrapped Newtonian gravity.
The starting point is the complete spacetime metric~\eqref{eq:g} derived in Ref.~\cite{Casadio:2021gdf}.
The leading order deviation from the Schwarzschild solution cannot be eliminated and is encoded in the free parameter
$\Xi$, which is not \textit{a priori} known and must be constrained by observations.
\par
First, we show that bounds on $\Xi$ can be deduced from the comparison between the measurements
of the orbital precession of Solar System bodies and the theoretical predictions arising from bootstrapped
Newtonian metric computed in Ref.~\cite{Casadio:2021gdf}.
The inferred confidence region for $\Xi$ for each planet is reported in Table~\ref{tab:tab1} and graphically
depicted in Fig.~\ref{fig:fig1}.
Based on the tightest interval obtained with Venus, we found that $\Xi$ lies in the range $[-1149.67\,;\,+1167.47]$.
With these values of the parameter $\Xi$ we predicted the orbital precession for Uranus, Neptune and Pluto,
and we found a theoretical precession in great agreement with the General Relativistic value.
Such a compatibility was confirmed by turning our attention to the Galactic Center and repeating the same
analysis for the star $S2$~\cite{GRAVITY:2020gka}.
The mean value of the parameter $\Xi$ such that the bootstrapped Newtonian precession equals the Schwarzschild
value is 
\begin{equation}
\Xi=-1.64236\pm0.10305
\ .
\end{equation}
\par
We next focused on the Galactic Center scale to constrain $\Xi$ by investigating the orbital motion of S-stars.
We used a fully relativistic approach based on an agnostic method:
for each value of $\Xi$, we solved the geodesic equations numerically starting from initial conditions at the
apocenter.
After applying the Thiele-Innes formulas to the mock positions, we were able to compare the resulting solution
with the observed stellar orbits.
Finally, we quantified the discrepancy between the simulated and observed orbits performing a $\chi^2$-statistics.
The inferred confidence region for $\Xi$ is compatible with the bounds obtained by the precession analysis,
and thus with General Relativity.
Indeed we found $17400_{-32244.3}^{+30555.6}$.
Since S-stars are at a distance of about  $r>1000\, R_g$ from the source, strong-field effects are not relevant,
and such a result was expected. 
\par
The proposed approach is completely general and represents a useful tool in the classification of extended
theories of gravity.
Moreover, this approach has already been used to test a Yukawa-like gravitational potential by means of dynamical tests at the
Galactic Center~\cite{yuk1,yuk2,yuk3,DellaMonica2021}, where no significant deviations from General Relativity were found.
Nevertheless, the definitive confirmation/exclusion of a given extended theory of gravity requires the improvement
of the constraints on its free parameters based on the observation of various strong-field effects.
This task can be accomplished taking advantage of the increasing high accuracy observations of second
generation instruments like GRAVITY~\cite{Gillessen:2010ei}.
\par
In particular, we focus on finding stars with short semimajor axis and highly eccentric orbits within the
pericenter of $S2$.
The existence of such a population of stars can be inferred from the recent discovery of the sources $S62$,
$S4711$ and $S4714$~\cite{peissker2020,peissker20202}.
Observing stars at smaller radii is essential to detect strong-field effects, which become no longer negligible
for distances of the pericenter $r\simeq 10\,R_g$, and therefore any deviations from General Relativity
to find out the underlying gravitational theory.

\begin{acknowledgments}
R.C.~is partially supported by the INFN grant FLAG. M.D.L. and A.D. acknowledges INFN Sez. di Napoli (Iniziativa Specifica TEONGRAV). 
A.G.~is supported by the European Union's Horizon 2020 research and innovation programme under the 
Marie Sk\l{}odowska-Curie Actions (grant agreement No.~895648--CosmoDEC).
The work of R.C.~and A.G~has also been carried out in the framework
of activities of the National Group of Mathematical Physics (GNFM, INdAM). 
\end{acknowledgments}

\appendix

\
\


\begin{thebibliography}{99}
%
%
%
\bibitem{HE}
S.~W.~Hawking and G.~F.~R.~Ellis,
``The Large Scale Structure of Space-Time,''
(Cambridge University Press, Cambridge, 1973)
%
\bibitem{geroch}
R.~P.~Geroch and J.~H.~Traschen,
Phys.\ Rev.\ D {\bf 36} (1987) 1017
[Conf.\ Proc.\ C {\bf 861214} (1986) 138];
%
%
\bibitem{BootN}
R.~Casadio, M.~Lenzi and O.~Micu,
  Phys.\ Rev.\ D {\bf 98} (2018) 104016
  [arXiv:1806.07639 [gr-qc]].
  %
\bibitem{Casadio:2020mch}
R.~Casadio and I.~Kuntz,
Eur. Phys. J. C \textbf{80} (2020) 581
[arXiv:2003.03579 [gr-qc]].
%
\bibitem{weinberg}
S.~Weinberg,
``Gravitation and Cosmology: Principles and Applications of the General Theory of Relativity,''
(Wiley \& Sons, 1972)
%
\bibitem{deser}
S.~Deser,
Gen.\ Rel.\ Grav.\  {\bf 1} (1970) 9
[gr-qc/0411023];
Gen.\ Rel.\ Grav.\  {\bf 42} (2010) 641
[arXiv:0910.2975 [gr-qc]].
%
\bibitem{Feynman}
R.~P.~Feynman, F.~B.~Morinigo, W.~G.~Wagner and B.~Hatfield,
``Feynman lectures on gravitation,''
(Addison-Wesley Publishing Company, Reading, 1995)
%
\bibitem{rubio}
R.~M.~Wald,
Phys.\ Rev.\ D {\bf 33} (1986) 3613;
K.~Heiderich and W.~Unruh,
Phys.\ Rev.\ D {\bf 38} (1988) 490;
 M.~P.~Hertzberg,
JHEP {\bf 1709} (2017) 119
[arXiv:1702.07720 [hep-th]];
D.~Bai and Y.~H.~Xing,
  Nucl.\ Phys.\ B {\bf 932} (2018) 15
  [arXiv:1610.00241 [hep-th]];
  R.~Carballo-Rubio, F.~Di Filippo and N.~Moynihan,
  JCAP {\bf 1910} (2019) 030
  [arXiv:1811.08192 [hep-th]];
%
  D.~Hansen, J.~Hartong and N.~A.~Obers,
  Phys.\ Rev.\ Lett.\  {\bf 122} (2019)  061106
  [arXiv:1807.04765 [hep-th]].
%
\bibitem{Casadio:2019cux}
  R.~Casadio, M.~Lenzi and O.~Micu,
   Eur.\ Phys.\ J.\ C {\bf 79} (2019) 894
  [arXiv:1904.06752 [gr-qc]].
%
\bibitem{Casadio:2020kbc}
R.~Casadio and O.~Micu,
Phys. Rev. D \textbf{102} (2020) 104058
[arXiv:2005.09378 [gr-qc]].
%
\bibitem{Casadio:2019pli}
R.~Casadio, O.~Micu and J.~Mureika,
Mod. Phys. Lett. A \textbf{35} (2020) 2050172
[arXiv:1910.03243 [gr-qc]].
%
\bibitem{Casadio:2016zpl}
R.~Casadio, A.~Giugno and A.~Giusti,
Phys.\ Lett.\ B {\bf 763} (2016) 337
[arXiv:1606.04744 [gr-qc]]
%
\bibitem{Casadio:2017cdv} 
R.~Casadio, A.~Giugno, A.~Giusti and M.~Lenzi,
Phys.\ Rev.\ D {\bf 96} 044010 (2017)
[arXiv:1702.05918 [gr-qc]].
%
\bibitem{ciarfella}
R.~Casadio, M.~Lenzi and A.~Ciarfella,
Phys. Rev. D \textbf{101} (2020) 124032
[arXiv:2002.00221 [gr-qc]].
%
\bibitem{DvaliGomez} 
G.~Dvali and C.~Gomez,
Fortsch.\ Phys.\  {\bf 61} (2013) 742
[arXiv:1112.3359 [hep-th]];
G.~Dvali, C.~Gomez and S.~Mukhanov,
``Black Hole Masses are Quantized,''
arXiv:1106.5894 [hep-ph].
G.~Dvali and C.~Gomez,
Phys.\ Lett.\ B {\bf 719} (2013) 419
[arXiv:1203.6575 [hep-th]];
Phys.\ Lett.\ B {\bf 716} (2012) 240
[arXiv:1203.3372 [hep-th]];
Eur.\ Phys.\ J.\ C {\bf 74} (2014) 2752
[arXiv:1207.4059 [hep-th]];
 A.~Giusti,
 Int.\ J.\ Geom.\ Meth.\ Mod.\ Phys.\  {\bf 16} (2019) 1930001.
%
\bibitem{Casadio:2021gdf}
R.~Casadio, A.~Giusti, I.~Kuntz and G.~Neri,
Phys. Rev. D \textbf{103} (2021) 064001
[arXiv:2101.12471 [gr-qc]].
%
\bibitem{cosmo}
M.~Cadoni, R.~Casadio, A.~Giusti, W.~M\"uck and M.~Tuveri,
Phys. Lett. B \textbf{776} (2018) 242
[arXiv:1707.09945 [gr-qc]].
%
%
\bibitem{DeMartino2018}I. De Martino, R. Lazkoz, M. De Laurentis Phys. Rev. D \textbf{97}, 104067 (2018)
%
\bibitem{Will2018}C. Will, Phys. Rev. Lett. \textbf{6120}, 191101 (2018).
%
\bibitem{Mayer1971}T. D. Moyer, Mathematical formulation of the Double-Precision Orbit Determination Program (DPODP)., Technical Report 32-1527 (NASA Jet Propulsion Laboratory, Pasadena, 1971).

\bibitem{Moyer200} T. D. Moyer, Formulation for observed and computed values of Deep Space Network data types for navigation, JPL Publication 00-7 (NASA Jet Propulsion Laboratory,Pasadena, 2000).

\bibitem{Eckart1996}A. Eckart and R. Genzel, Nature \textbf{383}, 415 (1996).
%
\bibitem{Ghez:2008}A. M. Ghez, S. Salim, N. N. Weinberg, J. R. Lu, T. Do, J. K. Dunn, K. Matthews, M. R. Morris, S. Yelda, E. E. Becklin, T. Kremenek, M. Milosavljevic, and J. Naiman, Astrophys. J.
\textbf{689}, 1044 (2008), arXiv:0808.2870.
%
\bibitem{Eckart1997}A. Eckart and R. Genzel, Mon. Not. R. Astron. Soc. 284, 576
(1997).
%
\bibitem{Gillessen:2009} S. Gillessen, F. Eisenhauer, S. Trippe, T. Alexander, R. Gen-
zel, F. Martins, and T. Ott, Astrophys. J. 692, 1075 (2009),
arXiv:0810.4674.
%
\bibitem{Gillessen2009L}S. Gillessen, F. Eisenhauer, T. K. Fritz, H. Bartko, K. Dodds-Eden, O. Pfuhl, T. Ott, and R. Genzel, Astrophys. J. Lett. 707, L114 (2009), arXiv:0910.3069 [astro-ph.GA].
%
\bibitem{Ghez1998} A. M. Ghez, B. L. Klein, M. Morris, and E. E. Becklin, Astrophys. J. 509, 678 (1998), astro-ph/9807210.



\bibitem{adm}
R.L.~Arnowitt, S.~Deser and C.W.~Misner,
Phys.\ Rev.\  {\bf 116} (1959) 1322.
 %
\bibitem{Gillessen:2017}
S.~Gillessen, PM.~ Plewa, F.~Eisenhauer, R.~Sari, I.~Waisberg, M.~Habibi, O.~Pfuhl, E.~George, J.~Dexter, S.~von Fellenberg, \textit{et al.}
Astrophys. J. \textbf{837} (2017) 30
[arXiv:1611.09144 [astro-ph.GA]].
%
\bibitem{Schodel:2003gy}
R.~Schodel, T.~Ott, R.~Genzel, A.~Eckart, N.~Mouawad and T.~Alexander,
Astrophys. J. \textbf{596} (2003) 1015
[arXiv:astro-ph/0306214 [astro-ph]].

\bibitem{Gillessen:2009ht}
S.~Gillessen, F.~Eisenhauer, T.~K.~Fritz, H.~Bartko, K.~Dodds-Eden, O.~Pfuhl, T.~Ott and R.~Genzel,
Astrophys. J. Lett. \textbf{707} (2009), L114
[arXiv:0910.3069 [astro-ph.GA]].

\bibitem{GRAVITY:2020gka}
R.~Abuter \textit{et al.} [GRAVITY],
Astron. Astrophys. \textbf{636} (2020) L5
[arXiv:2004.07187 [astro-ph.GA]].

\bibitem{1930MNRAS}
W.M.~Smart
Mon. Not. Roy. Astron. Soc. \textbf{90} (1930) 534.

\bibitem{aitken}
R.G.~Aitken
``The Binary Stars,''
(1964)

\bibitem{2015MNRAS.451.3034N}
G.G.~ Nyambuya, 
Mon. Not. Roy. Astron. Soc. \textbf{451} (2015) 3034.

\bibitem{2013MNRAS.432.3431P}
 E.V.~Pitjeva,  N.P.~ Pitjev
 Mon. Not. Roy. Astron. Soc. \textbf{432} (2013) 3431.

\bibitem{yuk1}
I.~de Martino, R.~della Monica and M.~de Laurentis,
``$f(R)$-gravity after the detection of the orbital precession of the S2 star around the Galactic centre massive black hole,''
[arXiv:2106.06821 [gr-qc]].
\bibitem{yuk2}
A.~D'Addio
Phys. Dark Univ. \textbf{33} (2021) 100871.
\bibitem{yuk3}
M. De Laurentis, I. De Martino, R. Lazkoz, Phys. Rev. D \textbf{97}, 104068 (2018).
%
\bibitem{DellaMonica2021}
R. Della Monica, I. de Martino, M. De Laurentis, arXiv:2105.12687 [gr-qc]
%
\bibitem{Gillessen:2010ei}
S.~Gillessen, F.~Eisenhauer, G.~Perrin, W.~Brandner, C.~Straubmeier, K.~Perraut, A.~Amorim, M.~Sch\"oller, C.~Araujo-Hauck and H.~Bartko, \textit{et al.}
Proc. SPIE Int. Soc. Opt. Eng. \textbf{7734} (2010) 77340Y
[arXiv:1007.1612 [astro-ph.IM]].

\bibitem{peissker2020}
F.~Peissker, A.~Eckart and M.~Parsa
Astrophys. J. \textbf{889} (2020) 61
[arXiv:2002.02341[astro-ph.GA]].

\bibitem{peissker20202}
F.~Peissker, A.~Eckart, M.~Zajacek, B.~Ali and M.~Parsa
Astrophys. J. \textbf{899} (2020) 50.





\end{thebibliography}
\end{document}